# Efficient and affordable thermoelectric measurement setup using Arduino and LabVIEW for education and research

Alex J. Oh[1,2], Colby J. Stoddard[1,3], Craig Queenan[2], and Seongshik Oh[1,3]

[1]Department of Physics and Astronomy, Rutgers, The State University of New Jersey, Piscataway, NJ 08854, USA

[2]High Technology High School, Lincroft, NJ 07738, USA

[3]Center for Quantum Materials Synthesis, Rutgers, The State University of New Jersey, Piscataway, NJ 08854, USA

## Abstract

Thermoelectric materials can convert thermal energy into electricity, making them promising candidates for harvesting waste heat, an increasingly important challenge in the energy-intensive modern world. The search for improved thermoelectric materials is therefore an active area of research in materials physics. Despite their fundamental and practical significance, thermoelectric properties—such as the Seebeck coefficient and power factor—are rarely explored in student labs due to the complexity in measurement schemes and requirement for sophisticated equipment. In this work, we present a user-friendly, low-cost and efficient thermoelectric measurement system built with Arduino and LabVIEW, which can simultaneously measure Seebeck coefficients and power factors as a function of temperature. This was made possible by improving the resolution of Arduino over ~1000 times with amplifiers and noise reduction schemes. With a total cost of only ~$100 and simple measurement protocols, this setup is well suited not only for student labs but also for efficient thermoelectric research.

## I. INTRODUCTION

Thermal energy is the waste product of almost all processes on Earth. One good example is the huge amount of heat generated by large data centers managing all internet activities around the globe. Therefore, a way to efficiently capture waste heat and turn it back into usable energy, which in modern society is mainly electricity, can be a significant development. There are already ways to convert heat to usable energy, such as steam generators in power plants. However, these types of energy converters are impractical for capturing waste heat. An efficient solution is using thermoelectrics, which are materials that can create a substantial electrical voltage from a temperature difference. They can convert waste heat into electricity without any moving parts and make extremely lightweight energy conversion devices.[1]

When a temperature difference is created within any material, a voltage forms due to the contrasting charge concentrations on the colder and hotter sides. Charge on the hotter side will have greater kinetic energy and will spread out, resulting in reduced charge density, and it will be opposite on the colder side. Such a charge imbalance leads to a voltage across the material, which can then be utilized as a source of electricity. This is known as the Seebeck effect, the main principle of thermoelectrics.[2] The opposite effect, called the Peltier effect, can be used to create a temperature difference from electricity. Devices that use the Seebeck effect to generate electricity are called thermoelectric generators, and devices that use the Peltier effect to cool are called thermoelectric coolers.[1,3]

The thermoelectric power, or Seebeck coefficient, is a property that determines the relationship between the temperature difference and the voltage in a material. It is a major factor that contributes to the quality of the thermoelectric material. Expressed in an equation form, the Seebeck coefficient $S$ is defined as $-\frac{\Delta V}{\Delta T}$, where $\Delta V$ is the voltage in (commonly) μV and $\Delta T$ is the temperature difference in K.[2] A high Seebeck coefficient means that the voltage generated by the material is large for a given temperature difference. It has been observed that better conductors typically have a lower Seebeck coefficient.[4] However, the overall quality of a thermoelectric material is usually determined by the dimensionless figure of merit, $ZT = \frac{S^2\sigma}{\kappa}T$, where $T$ is the material's temperature (in K), $S$ is the Seebeck coefficient, $\sigma$ is the electrical conductivity, and $\kappa$ is the thermal conductivity. Practically, the power factor, $S^2\sigma$, is used as a simpler measure of performance because thermal conductivity is much more difficult to measure, particularly for thin film materials.[1]

Despite the conceptual simplicity of the thermoelectric properties and their wide uses in everyday applications such as thermocouples and thermoelectric coolers, measurement of thermoelectric properties usually requires sophisticated instruments and complex measurement geometries that cannot be easily incorporated into student labs.[5–8] Recently, however, with the advent of low-cost microcontroller boards like Arduino, many advanced measurements, which were previously considered too complex to implement for a student lab, have become accessible with relative ease. Nonetheless, there is no report of utilizing Arduino to measure thermoelectric properties in the literature.

Here, we present an efficient thermoelectric measurement setup (Figs. 1 and 2), simultaneously measuring both Seebeck coefficient and resistance, thus power factor, with an Arduino Uno board and a LabVIEW (Laboratory Virtual Instrument Engineering Workbench) user interface. This setup was verified on a bare Chromel wire (0.002" diameter, from Omega), which has well-documented Seebeck coefficients and resistivity values in the literature.[2,9] Subsequently, it was used to measure a $Bi_2Se_3$ thin film (30 nm thick, grown on a $10 \times 10 \times 0.5$ mm$^3$ $Al_2O_3$ (0001) substrate) over a range of temperatures and confirmed that the measured values are consistent with those in the literature,[10] demonstrating that this setup can also be used for thermoelectric materials research.

In order to use Arduino for thermoelectric measurement, several challenges need to be overcome. First, we need to address the poor resolution of the Arduino Uno board. Because Arduino is equipped with a 10-bit digitizer and operates on the 5 V scale, its resolution is $5 \text{ V} / 2^{10} = \sim 5 \text{ mV}$,[11] which is too coarse to measure thermoelectric voltages, which can be well below ~1 mV. Second, for temperature-dependent thermoelectric properties, we need to precisely control the temperature and its gradient across the sample while measuring the small thermoelectric signals. Third, we need to find a way to measure electrical resistance as well as the Seebeck coefficient for the power factor. Finally, for this setup to be effectively utilized for student labs, we need to find a simple way to make electrical contacts to the samples. With the setup described below, we have successfully addressed all these challenges.

As for the software portion of the setup to control Arduino, although it is most common to use an interface with the C/C++-based Arduino language, we chose LabVIEW instead because it makes it much easier to collect and organize data in real time with a custom graphic user interface (GUI). Due to these many benefits that LabVIEW provides, there have been continued efforts in the literature trying to combine LabVIEW with Arduino for various applications. [12–14] Nonetheless, efforts related to thermoelectric measurements are still lacking.

Except for the $Bi_2Se_3$ thin film, which was synthesized in our lab, all the other parts used in this setup can be purchased online at affordable prices: see Table A in the Appendix for the information on the parts, including their prices and vendors. Another notable point, as detailed below, is that the way electrical contacts are made is greatly simplified in this setup in comparison to the common practices in research labs.[5–7,15–17] So, the unique and simple sample mounting schemes introduced here can be useful even for frontier research labs.

## II. EXPERIMENTAL SETUP

When we set out with this project, our goal was to build a user-friendly thermoelectric setup using only Arduino and easily available accessories, so that it can be understood, built, and tested by someone with only a basic electronics background. So, when there were ready-made Arduino-compatible electronic modules available that could reduce the burden of design on us, we decided to use them rather than try to build the equivalent circuit on our own. With this basic design strategy, we provide details of the experimental setup and the design process below, including discussions on how and why we took certain components or approaches rather than others. We hope that with the provided details, the readers -- students, teachers, or individual researchers -- will be able to further improve or modify the setup to their needs.

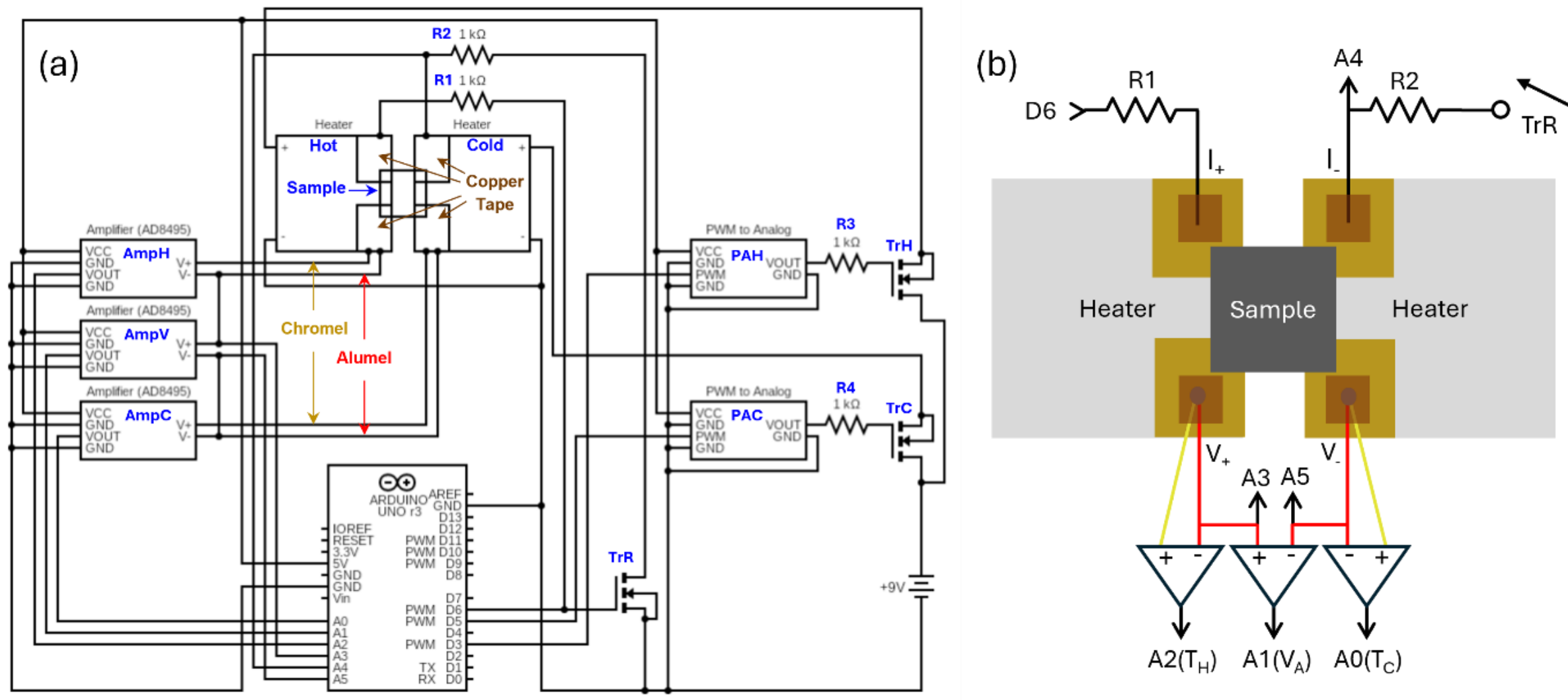

**FIG. 1. (Color online) Circuit diagrams of the setup: (a) Entire setup and (b) Simplified diagram, focusing on the sample area**. In (b), Chromel and Alumel wires are represented as

yellow and red, respectively, following the industry convention of their insulations. Heat sinks for the transistors and heaters were also used but are not shown here.

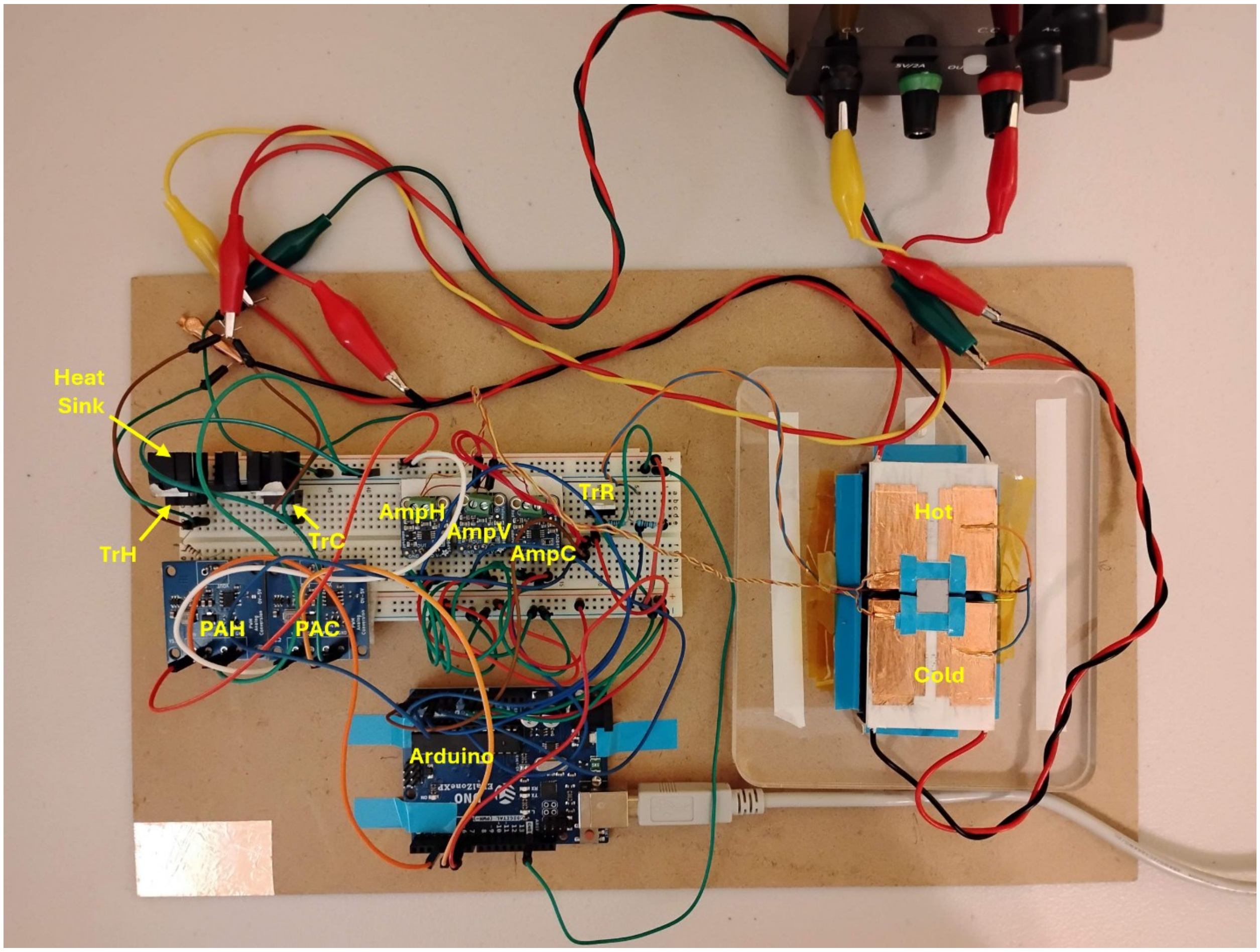

**FIG. 2. (Color online) Photo of the setup.** Note that wires with sensitive signals such as the thermocouple wires, or with large currents (for the heaters) are twisted in pairs to minimize inductive noise pickup and noise generation, respectively. Key components are also labeled for easy comparison with the circuit diagram in Fig. 1.

## A. Temperature measurement and control

The first step of the project is to measure and control the temperatures and their gradients across the sample. For this purpose, two thermoelectric heaters/coolers (model TEC1-12706, lateral size of 40 × 40 mm$^2$) were used to heat the two sides of the sample (see Figs. 1 and 2). In this study, we used them as heaters, but they can also be used as coolers to study thermoelectric properties below room temperature. These heaters are flat plates that use thermoelectrics to heat one face and cool the other when electric current is supplied; thus, by reversing the current flow direction, these devices can be converted to coolers; they can cool or heat from below 0 ˚C to ~100 ˚C. We have chosen these thermoelectric heaters mainly because they provide level surfaces with both heating and cooling options. Here, we tested the system only above room temperature with the heating option, but it should be possible to operate it below room temperature using the cooling option just by reversing the current flow direction of each thermoelectric heater/cooler.

A general-purpose power supply that can provide more than 10 V and 3 A was used to power the heaters, and two conducting aluminum Easycargo heat sinks (model 404010B10-8810) were used to equalize the temperature of the cooling side of the heaters to the ambient. K-type insulated thermocouple wires that contain Chromel and Alumel with AWG 30 (0.254 mm diameter) from Minnesota Measurement Instruments were used for the temperature measurement. Copper tape (Bomei Pack) held the wires together on the heaters. Three Adafruit Analog Output K-Type thermocouple amplifiers (model AD8495) measured and amplified the voltage signals from the thermocouple wires.

To control and automate the temperature of the heaters, a Proportional-Integral-Derivative (PID) controller was used for each heater. A PID controller is a commonly used negative feedback system that can keep a measured value close to a target value.[13] An example of such a controller is a home temperature regulator. For the hardware portion of the PID controller, two N-channel

MOSFET transistors (model RFP30N06LE) were first placed on the breadboard (TrH and TrC). The software portion is described below in Sec. IID. Then, a heat sink, as shown in Fig. 2, was attached to the two transistors, electrically isolated from each other, in order to dissipate heat during operation: we used a single heatsink just because that was what we had in the lab, but two separate heat sinks would have been better to minimize coupling between the two transistors. These transistors were configured in the source-follower mode, so the positive terminal of the power supply was directly connected to the drain of the transistors. Then, the sources were connected to the heaters in such a way that they would heat up (rather than cool down) the sample when the transistors (TrH and TrC) were turned on. In this mode, the voltage at the source follows the voltage at the gate (minus a small threshold voltage), as long as the voltage at the gate is less than the voltage at the drain but higher than the threshold voltage. Next, the remaining wire of each heater was connected to the negative terminal of the power supply and the digital ground of the Arduino.

We chose this source-follower mode instead of connecting the heater on the drain side, because the PID control worked much more stably with the almost linear response of the source-follower mode: a non-linear response does not work effectively with the PID control. Considering that in the source-follower mode, the amplification between the input and output voltages is effectively one, the only critical factor when choosing the right transistor is its capacity to dissipate large heat: in our case, it should be able to dissipate a maximum of 10 V and 1 A, or 10 W. That is also why we attached the transistors to a heatsink to effectively dissipate the heat away from the transistors.

Then, two Pulse Width Modulation, or PWM outputs (D3 and D5) from the Arduino were each connected to EC Buying PWM-to-analog converters (PAH and PAC) on the breadboard, the outputs of which were connected to the gate pins of the transistors. PWM simulates analog outputs through digital signals by rapidly alternating between 0 and 5 V such that the average matches the desired analog voltage between 0 and 5 V. The converters were used because using raw PWM signals to power the heaters resulted in substantial noise in Seebeck coefficient measurements due to electromagnetic induction caused by the alternating current intrinsic to the PWM mode. Furthermore, the PWM-to-analog converter had the option of converting the PWM signal to 0 – 10 V (instead of 0 – 5 V) analog signal, and this allowed us to control the voltage across the heaters up to 10 V of the power supply voltage; 5 V of power supply voltage was insufficient to heat the sample up to 100 ˚C. The two resistors (R3 and R4) were used as part of low-pass *RC* filters -- instead of the PWM-to-analog converters -- in an earlier version of the setup. A low-pass *RC* filter with a large capacitor (10 μF) connected between the resistor and the ground can work almost like a PWM-to-analog converter, but due to the output voltage being limited to 5 V, we could not reach 100 ˚C with the RC filter. So, in the end, we chose the commercially available PWM-to-analog converter, and R3 and R4 can be removed without any negative effect. Nonetheless, we kept them just because they seem to help reduce inductive noises.

**B. Seebeck coefficient measurement**

In order to measure and control temperatures and voltages across the sample, patches of copper tapes were used at the four corners of the sample (each, 3 mm × 3 mm of an area) on top of the heaters as shown in Figs. 1 and 2. For Seebeck measurements, only the bottom two contact points in the sample area of Fig. 1 (the left two contacts of the sample area in Fig. 2) were used. Unless stated otherwise, all the position or orientation references below will be based on Fig. 1 instead of Fig. 2: the heater and sample configurations in Fig. 2 are rotated 90˚ from those in Fig.

1. The top two contacts were used as current input for electrical resistance measurements, which will be explained below in Sec. IIC. Small pieces of thermally conducting, yet electrically insulating, tape (blue tape in Fig. 2) were put on top of the copper tape and carefully cut to form the border of a square to accept the 10 mm × 10 mm sample with a small bit of copper tape at the corners. This is where the samples were placed, face down. The main purpose of the blue tape is to define the sample area, and we used thermally conducting tape in order to minimize temperature nonuniformity on each side of the two heaters.

One Chromel and one Alumel wire (AWG 30, each with yellow and red insulation, respectively) were spot-welded together at one end with the insulations removed, forming a K-type thermocouple, and this welded side was attached with copper tape to a corner of the sample on the hotter side [see Figs. 1(b) and 2]: those who do not have access to a spot-welder can purchase a pre-welded thermocouple (say, from Minnesota measurement instruments or Omega) or may just twist the wires without much compromising the performance due to the surrounding copper tape. The other ends of each thermocouple were connected to an AD8495 thermocouple amplifier on the breadboard. The same was done on the colder side. The two AD8495 amplifiers (AmpH and AmpC) acted as thermometers and measured the temperatures ($T_H$ and $T_C$) of the two sides of the sample, with the temperature difference $\Delta T = T_H - T_C$ being used for the Seebeck coefficient calculation. A third amplifier (AmpV) was connected to the two Alumel wires of the thermocouples with two more Alumel wires, and this measured the voltage ($\Delta V$) across the sample. Here, we used the Alumel wires of the temperature measurement thermocouples as part of the voltage probing wires in order to simplify the wiring. However, one can instead use a dedicated pair of Alumel wires all the way from the sample to the voltage amplifier (AmpV). Interested readers can try this other approach to see if it improves the performance.

We chose the AD8495 breakout board from Adafruit because it allows Arduino to amplify voltages from a K-type thermocouple and convert them to temperatures without any additional circuitry. That is also why we chose the K-type thermocouple to measure the temperatures and voltages in our setup. However, there are other chips, such as MAX31855, that allow measurement of the temperatures with higher precision, using a digital, instead analog, interface. Interested readers can try these other options and see if they improve the performance.

The Seebeck coefficient $S$ can be calculated using the formula $S = -\frac{\Delta V}{\Delta T}$.[2] However, it is important to note that this is not the Seebeck coefficient of the sample $S_S$, because we have to consider the effect of the Alumel probing wire, i.e., $S_{Al}$, as follows. If we take into account the thermoelectric voltage developed across the Alumel wires, the voltage $\Delta V$ measured between the positive (+) and negative (-) ports of the amplifier (AmpV), is given by $\Delta V = -[\int_{T-}^{TC} S_{Al}(T)dT + \int_{TC}^{TH} S_S(T)dT + \int_{TH}^{T+} S_{Al}(T)dT]$.[2,4] If we rightfully assume that the temperatures $T$- and $T$+ at the ports of AmpV are the same, this equation reduces to $\Delta V = -\int_{TC}^{TH}[S_S(T) - S_{Al}(T)]dT \equiv -\int_{TC}^{TH} S(T)\,dT$. In other words, the Seebeck coefficient of the sample can be found by adding the contribution of the probing wires to the measured value: $S_S = S + S_{Al}$. Here, we can equally well choose Chromel as the probing wire; then, $S_{Al}$ should just be replaced by $S_{Ch}$.

The outputs of the three AD8495 amplifiers [AmpC, AmpV and AmpH in Fig. 1(a)] were connected to the input pins A0, A1, and A2 of the Arduino board (measuring $T_C$, $\Delta V$, and $T_H$, respectively) using jumper wires, and the amplifiers were also connected to the 5 V and ground. These amplifiers are used to measure both the temperatures and the thermoelectric voltages across the sample. The amplifier multiplies the input voltage by 122.4 and adds an offset, determined by the ambient temperature, according to the data sheet of AD8495.[18] Using the amplified voltage output ($V_T$) from the temperature-measuring amplifiers (AmpC and AmpH), each temperature ($T_H$ and $T_C$) can be calculated with the formula $T = \frac{V_T - 1.25}{0.005}$ °C.[18] We confirmed that the temperatures measured this way through the Arduino are consistent with the temperatures read by commercial K-type thermocouple readers. Considering that the resolution of the Arduino unit is ~5 mV and that the Seebeck coefficient of K-type thermocouple is about 41 μV/K, with the amplification of 122.4, AD8495 provides only 1 ˚C of temperature resolution in a single shot measurement.[18] However, by averaging over 100 data points, we are able to achieve a temperature resolution of 0.1 ˚C. With a temperature difference of 10 ˚C between the hot and cold temperatures, this method guarantees an uncertainty of 1% for the temperature measurement, which is sufficient for most purposes.

Although AD8495 is designed to measure temperatures of K-type thermocouples, we can also use this amplifier to measure the small thermoelectric voltages across the sample, with an amplification of 122.4. The only caveat is that its output voltage has an offset depending on the ambient temperature. Despite this issue, we figured that using the same amplifier would simplify the overall design rather than searching for and testing another amplifier to measure the thermoelectric voltage. The voltage across the sample can be obtained by $\Delta V = \frac{V_A - V_{ref}(T_{ref})}{122.4}$, where $V_A$ is the voltage measured at A1, amplified by AmpV, and $V_{ref}(T_{ref}) = 0.005T_{ref} + 1.25$ is the offset voltage, internally generated by AmpV, based on the ambient temperature $T_{ref}$.[18] $V_{ref}$ can be found by measuring $V_A$ when the sample temperature difference is zero, that is, when $T_C = T_H = 40$ ˚C, 50 ˚C, etc. Then, we subtract this offset voltage from $V_A$ to find $\Delta V$ at each sample temperature, as shown above, to take into account any drift in the offset voltage of the amplifier during the measurement.

### C. Electrical resistivity measurement

To create the setup for resistance measurement, a wire was first attached to each of the upper two copper patches of the sample area with copper tape as shown in Figs. 1 and 2, and these wires were connected to the breadboard. Together with the lower two contacts, which were also used for the Seebeck voltage measurement, these copper patches form the four-point van der Pauw geometry of the resistance measurement.[19] Next, on the breadboard, a 1 kΩ resistor (R1) was connected between the hotter side of the sample and a digital output (D6) of the Arduino that could provide a current through the sample. Then another 1 kΩ resistor (R2) was connected between the colder side of the sample to the drain of an N-channel MOSFET (TrR), with its gate being controlled by the same digital output (D6) providing the current: see Fig. 1(b). The source of the transistor was connected directly to the digital ground of the Arduino. This transistor (TrR), which disconnected the ground from the sample when the resistance was not being measured, was needed because the sample being directly connected to the ground interfered with the Seebeck measurement.

The digital output D6 is normally left in the floating mode, so that the current flow path of the resistance measurement is completely isolated from both the 5 V and the ground in order not to interfere with the Seebeck coefficient measurement. Then, when we want to measure the resistance, D6 is set to 5 V, which then turns on TrR and allows current to flow from 5 V → R1 → sample → R2 → TrR → ground. Because the (drain-to-source) voltage across TrR is negligible when turned on, the current $I$ flowing through the sample can be determined by measuring the voltage across R2 using A4: that is, $I$ = A4/R2. So, R2 provides a way to measure the electric current flowing through the sample during the resistance measurement. The addition of R1 between the 5 V line and the sample makes the electrical potential of the sample symmetrical with respect to the 5 V and ground lines; we found that such symmetrization helps reduce the inductive noise.

In order to choose the best resistance value for R1(R2), we need to first consider the maximum stable current that Arduino digital output can handle, which is ~20 mA according to the specification. With the digital output voltage of 5 V, this implies that the minimum resistance that should be put between D6 and ground should be 5 V / 20 mA = 250 Ω. This puts the lower bound for R1(R2) to be 125 Ω. Another factor to consider is the range of sample resistance to measure. Considering that the smallest voltage that Arduino can measure with a single shot measurement is 5 mV, while the applied voltage at D6 is 5 V, if we choose R1(R2) = 1 kΩ, the minimum sample resistance value we can measure directly with Arduino will be approximately (R1 + R2) × (5 mV / 5 V) = 2 Ω: this is because the sample forms a voltage divider with R1 and R2 in series and the minimum voltage that can be measured across the sample is 5 mV when the voltage across R1 + R2 is approximately 5 V. Similarly, the maximum measurable resistance is estimated when the voltage across R2 is 5 mV and the voltage across the sample is close to 5 V, which gives R2 × (5 V / 5 mV) = 1 MΩ. So, the measurable resistance range with R1(R2) = 1 kΩ would be approximately 2 Ω ~ 1 MΩ. If we choose instead R1(R2) = 1 MΩ, the measurable range will be 2 kΩ ~ 1 GΩ. Considering that the resistance of most thermoelectric materials is likely to be below 1 MΩ, R1(R2) = 1 kΩ is a reasonable choice.

Next, analog inputs A3 and A5 were connected to the Alumel wires of the thermocouples at the input of the voltage-measuring amplifier (AmpV): see Fig. 1(b). These inputs are used to measure the voltage $V = A3 - A5$ between the lower corners of the sample when D6 is set to 5 V. The resistance of the sample can then be calculated using the formula $R = V/I$. For low resistance (<~10 Ω) samples, because the voltage across the sample (A3 – A5) approaches the resolution of Arduino, the amplified signal ($V_A$) at A1 (taking into account the 122.4 amplification factor and the temperature-dependent offset as discussed in the previous section) can be used instead for more accurate resistance measurement. On the other hand, if the sample resistance is bigger than ~10 Ω, $V_A$ at A1 is saturated at a device maximum value (~4.5 V), and then, A3 – A5 should be used instead for the resistance measurement in such a case. This can be understood as follows. With the current through the sample being approximately, 5 V / 2 kΩ = 2.5 mA, sample resistance of 10 Ω results in 25 mV of voltage between A3 and A5. When this voltage is amplified by AmpV, with the amplification of 122.4 and the offset voltage of ~1.4 V added, the voltage approaches the system maximum of 5 V. Accordingly, if the resistance of sample is larger than ~10 Ω between the two voltage probes (A3 and A5), it cannot be measured with the AmpV amplifier.

Summarizing the above analyses regarding the resistance measurement, for samples more resistive than ~10 Ω (up to ~1 MΩ) between the voltage probes, A3 - A5 should be directly used to measure the resistance of the sample. If the sample resistance is less than ~2 Ω, the amplified signal $V_A$ should be used instead. If the sample resistance is between ~2 and ~10 Ω, both A3 – A5 and $V_A$ can be used to measure the resistance, but considering that $V_A$ provides much larger signals, using $V_A$ would provide higher precision of the resistance values. For the samples we tested, the measured resistance of the Chromel wire was about ~2.5 Ω, whereas that of the $Bi_2Se_3$ thin film was hundreds of Ohms, as shown below in Sec. IV. So, we used $V_A$ to measure the resistance of the Chromel wire and A3 – A5 for the $Bi_2Se_3$ film.

In order to convert the measured resistance values to resistivity, the geometry of the samples should be considered. For the wire sample, if the cross-sectional radius of the wire is $r$ and the length between the voltage probes is $l$, the resistivity, $\rho$, of the wire can be obtained from the measured resistance $R$ by $\rho = R \cdot \pi r^2 / l$. For the thin film sample, by measuring the resistances of two different orientations of the sample, one rotated 90˚ from the other, the sheet resistance $R_S$ can be found approximately by $R_{avg} \cdot \frac{\pi}{ln\ 2}$ according to the van der Pauw method.[19] From the sheet resistance, the resistivity can be calculated based on the thickness, $t$, of the sample by $\rho = R_S \cdot t$.

## D. Measurement software

Figure 3 shows the LabVIEW GUI we created to measure and control the setup. The GUI of the LabVIEW program displays and uses the analog inputs from the Arduino to automatically calculate and show the Seebeck coefficient, the temperatures, the voltage, and the resistance, both in graphs and in texts. The LabVIEW program measures all six analog channels (A0 through A5), every 0.5 s, at temperatures from 40 °C to 100 °C. In order to set the temperature to a specific value, we utilized the PID function provided by the LabVIEW driver with the hardware described in Sec. IIA: the optimal values of P, I, and D parameters were determined empirically.

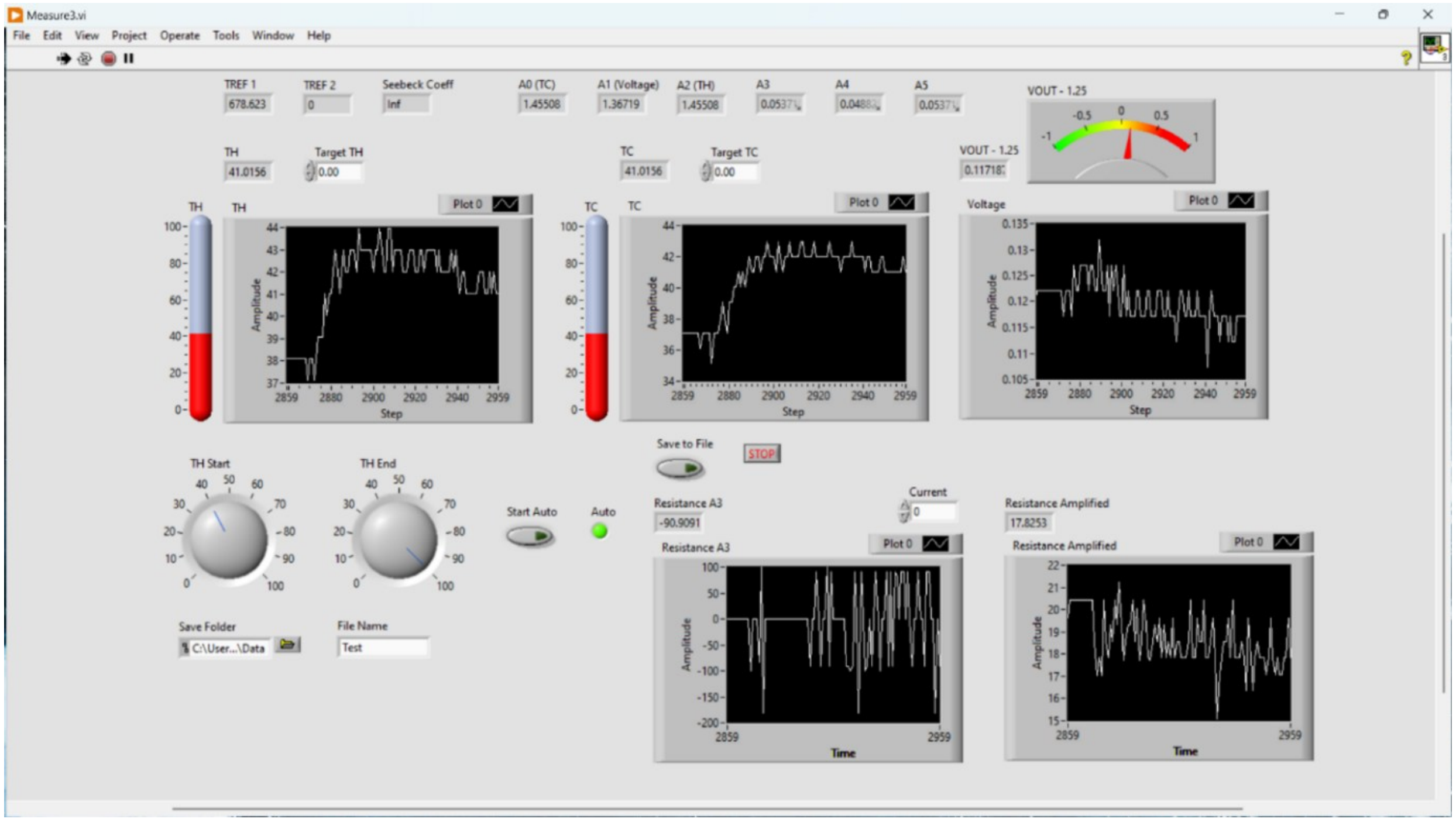

**FIG. 3. (Color online) GUI of the LabVIEW program.** This program automates all the controls and measurements as described in the main text. The full LabVIEW code is provided in supplementary material.

The program first set the temperatures of both heaters to 40 °C and started measuring the temperatures and voltage (A0 for $T_C$, A2 for $T_H$, and A1 for $\Delta V$). Then, current for the resistance measurement was supplied by applying 5 V through D6, and resistances were measured using the analog inputs on the Arduino: A4 for the current measurement, A3 and A5 for the unamplified voltage measurement, and A1 for the amplified voltage measurement.

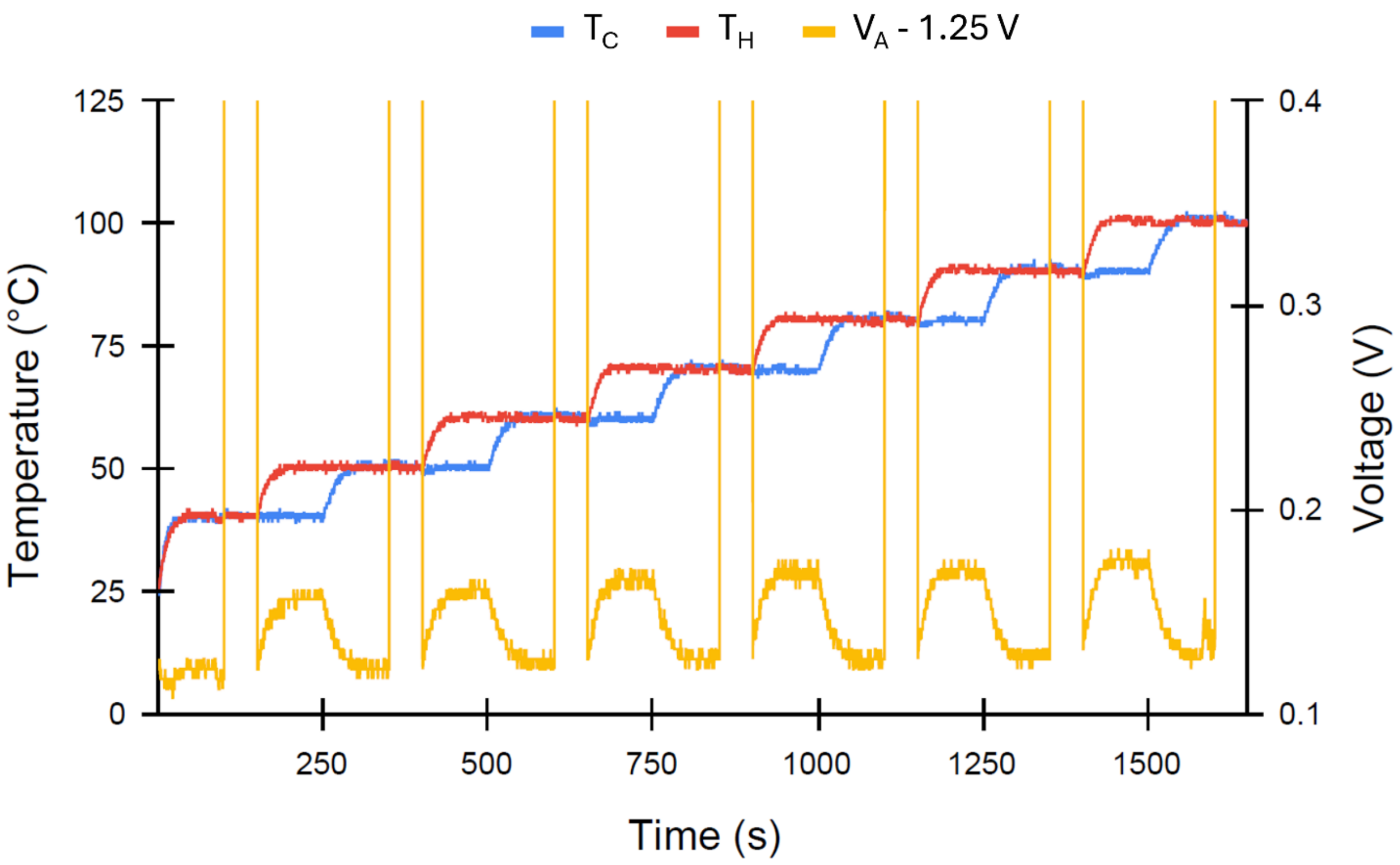

**FIG. 4. (Color online) Example time sequence of temperatures and voltage data.** This data shows a time sequence of the cold ($T_C$) and hot ($T_H$) temperatures - $T_C$ from A0 and $T_H$ from A2,

ranging from 40 ˚C through 100 ˚C - and the amplified voltage ($V_A$) at A1 – 1.25 V. Note that $V_A$ represents amplified Seebeck signals when $T_H - T_C = 10$ ˚C and amplified resistance voltages when $T_H - T_C = 0$. As discussed in Sec. IIC, if the sample resistance is bigger than ~10 Ω, the amplified resistance voltage reaches close to 5 V and saturates, due to the amplification and offset voltage of AmpV: the saturated voltages are outside the range in this plot. In such a case, A3 - A5 values should be used instead for the resistance measurement.

Next, the temperature $T_H$ of the one heater was increased by 10 ℃ to 50 ℃, while constantly recording the amplified thermoelectric (Seebeck) voltages at A1. During the Seebeck coefficient measurement, the current-flow path for the resistance measurement was disconnected from the circuitry in order not to interfere with the Seebeck coefficient measurement, by floating D6, which isolates the sample from 5 V as well as from the ground by turning off the transistor (TrR), as described in Sec. IIC. Then, the temperature $T_C$ of the other heater was increased to 50 ℃, and the offset voltage and resistance were measured again.

This cycle continued until the temperatures reached 100 ℃, at which point the program terminated and saved the data into a CSV file: see Fig. 4 for an example data sequence. During post-data-processing, 100 data points of each signal were averaged at each temperature and used as the value corresponding to the temperature. This process reduces the uncertainty of the data by ten times, and combined with the amplification factor of 122.4, improves the resolution of the Arduino unit by ~1000 times; it also provides the standard deviation as an estimate of the data uncertainty.

## III. MEASUREMENT PROCEDURE

Figure 5 shows how the measurement was done on the $Bi_2Se_3$ film. The sample was placed face down on the sample holder. A rubber foot was placed on the back of the sample, and a weight (~250 g) was placed on top to provide good electrical contact between the sample and the copper tapes. The power supply was then turned on, and the voltage output was set to 9 V. Next, the automated LabVIEW program shown in Fig. 3 was run and the data (A0 through A5) were recorded every 0.5 s, as the temperatures swept from 40 ˚C to 100 ˚C with 10 ˚C steps as shown in Fig. 4. After the program finished running, the sample was then rotated 90˚ and the program was run again. The measured Seebeck coefficients and resistances were averaged between the two orientations, and the sheet resistance was calculated following the van der Pauw scheme, as described in Sec. IIC.

It is notable that even though we used a $Bi_2Se_3$ thin film sample for the test here, any slab-geometry bulk samples could be equally well used for the setup. There are many companies that sell conducting or semiconducting substrates of such a geometry. For example, doped Si and variety of other substrates with $10 \times 10 \times 0.5$ mm$^3$ dimensions can be purchased from, say, MTI corporation at affordable prices, and can be equally well measured using this setup following an identical protocol.

For the measurement of the Chromel wire, we used the same setup by placing the wire on the four copper contacts in a U shape. The two end contacts were used for current input during resistance measurement, and the other two middle contacts were used for the temperature and voltage measurements for the Seebeck coefficient and resistance measurements as described in Secs. IIB and IIC. The rubber foot and weight were also placed on top of the wire sample to keep good electrical and thermal contacts. We tested the Chromel wire first to confirm the proper operation of the setup and then moved on to measure the $Bi_2Se_3$ film.

In this setup, it is notable that we have greatly simplified the way electrical contacts are made. For any transport measurements, making reliable electrical contacts is one of the most critical steps, requiring significant effort and sophistication.[4,7,16,17] Such complexity is another factor making it difficult to implement thermoelectric measurement in student labs. On the other hand, our method of making electrical contacts by simply using a weight to press the sample from above eliminates the complications involved in making good electrical contacts: such a simplified scheme will not only lower the barrier to incorporating thermoelectric measurements for inexperienced students but also help save efforts and time even for professional researchers.

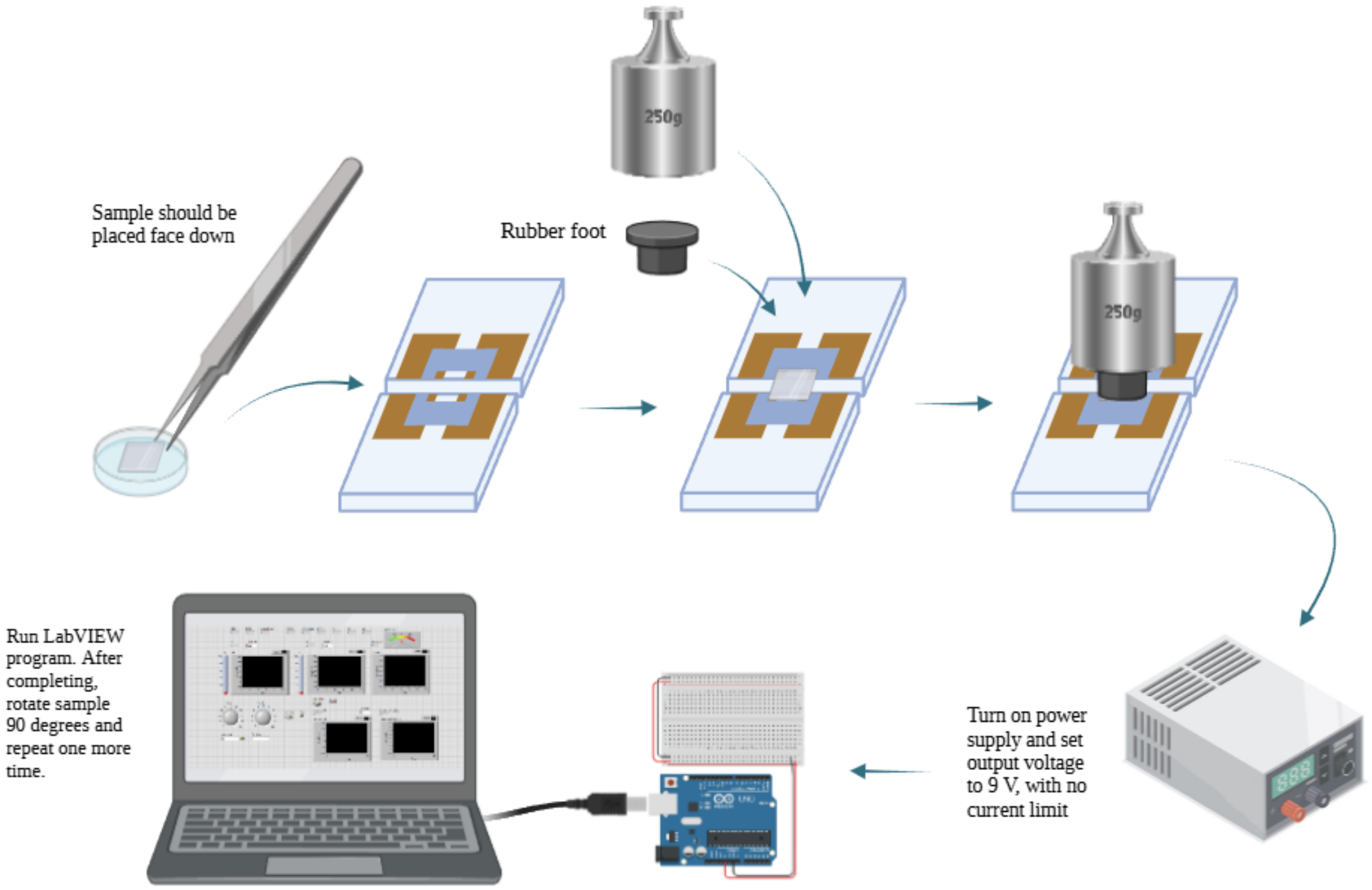

**FIG. 5. (Color online) Sample measurement procedure.** This diagram outlines the steps to measure a thin film sample on an insulating substrate. For a bulk sample, the term "face down" can be ignored.

## IV. TEST RESULTS AND DISCUSSION

Table I shows the Seebeck coefficients and the resistivities measured on the Chromel wire (0.002" in diameter) and compares them with the literature values in the temperature range from 40 °C to 100 °C.[2,9] As explained in Sec. IIB, because we used Alumel wires to measure the voltage between the hot and cold sides of the sample, we need to add the Seebeck coefficients of the Alumel wire to the measured values to obtain the Seebeck coefficients of the sample (here, Chromel wire). So, if $T_H$ is 50 °C and $T_C$ is 40 °C, we took the midpoint (45 °C) as the representative sample temperature, and added the known Seebeck coefficient $S_{Al_L}$ of the Alumel wire at 45 °C to the measured value $S_{K_M}$ to get the value $S_{Ch_M}$ of the Chromel wire: that is $S_{Ch_M} = S_{K_M} + S_{Al_L}$. Uncertainty in $S_{Ch_M}$ is estimated from the standard deviations of 100 data points at each temperature. One can see that $S_{Ch_M}$ values are all consistent with the literature values for $S_{Ch_L}$ within the standard deviations (~10%).

**TABLE I. Seebeck coefficient and resistivity measurement on a Chromel wire.** Literature values of K-type thermocouple ($S_{K_L}$), Alumel wire ($S_{Al_L}$), and Chromel wire ($S_{Ch_L}$) were obtained from Ref. 2. As for the resistivity, its literature values ($\rho_{Ch_L}$) are based on a document from Omega.[9]

| $T$ (°C) | $S_{K_M}$ (μV/K) | $S_{K_L}$ | $S_{Al_L}$ | $S_{Ch_M}$ | $S_{Ch_L}$ | $R_{Ch_M}$ (Ω) | $\rho_{Ch_M}$ (μΩ·m) | $\rho_{Ch_L}$ |
|---|---|---|---|---|---|---|---|---|
| 45 | 38.2 ± 3.4 | 41.1 | -18.5 | 19.7 ± 3.4 | 22.6 | 2.45 ± 0.02 | 0.69 ± 0.10 | 0.714 |
| 55 | 41.1 ± 3.4 | 41.3 | -18.4 | 22.7 ± 3.4 | 22.9 | 2.61 ± 0.02 | 0.73 ± 0.10 | 0.716 |
| 65 | 39.4 ± 3.1 | 41.5 | -18.3 | 21.1 ± 3.1 | 23.2 | 2.64 ± 0.03 | 0.74 ± 0.10 | 0.719 |
| 75 | 42.0 ± 3.6 | 41.6 | -18.1 | 23.9 ± 3.6 | 23.5 | 2.46 ± 0.03 | 0.69 ± 0.10 | 0.722 |
| 85 | 40.9 ± 4.2 | 41.5 | -17.9 | 23.0 ± 4.2 | 23.6 | 2.43 ± 0.02 | 0.68 ± 0.10 | 0.725 |
| 95 | 41.5 ± 3.1 | 41.4 | -17.7 | 23.8 ± 3.1 | 23.7 | 2.39 ± 0.04 | 0.67 ± 0.10 | 0.728 |

As for the resistivity measurement, we converted the measured resistance $R_{Ch_M}$ to the resistivity $\rho_{Ch_M}$ values using $\rho = R \cdot \pi r^2/l$, where $r$ = 0.001" (25 μm) and $l$ = 7 ± 1 mm. Note here that the biggest uncertainty for the resistivity estimation comes from that of the length estimation between the voltage probes. Because each copper tape probe contacting the Chromel wire has a contact area of 3 mm × 3 mm and the gap between the two probes is 4 mm, the center-to-center distance between the probes is 7 mm. Then, considering the irregular layout of the wire between the copper contacts and the relatively large probe size, we estimated the uncertainty in the length to be ±1 mm. Accordingly, even though the standard deviation of $R_{Ch_M}$ is only about a percent, the uncertainty of $\rho_{Ch_M}$ is much bigger (~13%) due to the length uncertainty. Within this uncertainty, $\rho_{Ch_M}$ of the Chromel wire is fully consistent with the literature values $\rho_{Ch_L}$, confirming the validity of the resistance measurement function of the setup.

Now that we have confirmed that the setup works as expected with the standard Chromel wire for both Seebeck coefficient and resistivity measurements, we report in Table II a full set of thermoelectric properties, including Seebeck coefficient, sheet resistance, conductivity, and power factor, on a 30-nm-thick $Bi_2Se_3$ thin film. Unlike the Chromel wire, which has well-defined Seebeck coefficients and resistivity values, the properties of $Bi_2Se_3$ thin films vary considerably depending on the exact growth details,[20,21] so there do not exist standard Seebeck coefficient values for $Bi_2Se_3$ thin films. With this in mind, we note that our Seebeck coefficient and sheet resistance values are about half of those reported by another group in the literature.[10] It is important to note that this does not imply that the accuracy of our setup can be that much off. Rather, while the accuracy of our setup is confirmed with the Chromel wire, such calibration information is missing in the previous study of $Bi_2Se_3$ thin films. So, there is a good chance that the previously reported measurement may be less accurate than ours. Second, the previous study used a $SrTiO_3$ substrate, whereas we used a $Al_2O_3$ substrate with different growth conditions, and it is well known that the transport properties of $Bi_2Se_3$ thin films depend strongly on the growth details. [20,21] With these factors taken into account, it is fair to say that these two results are reasonably consistent with each other.

**TABLE II. Thermoelectric property measurement on a $Bi_2Se_3$ thin film.** *S* represents the Seebeck coefficient of the $Bi_2Se_3$ film after taking into account the effect of the probing wire (Alumel), as explained in Table I and Sec. IIB.

| $T$ (°C) | $S$ (μV/K) | $R_S$ (Ω/□) | σ (kS/m) | $S^2\sigma$ ($\frac{\mu W}{m \cdot K^2}$) |
|---|---|---|---|---|
| 45 | -47.0 ± 2.0 | 347 ± 10 | 96.0 ± 2.9 | 212 ± 19 |
| 55 | -47.7 ± 2.1 | 361 ± 10 | 92.4 ± 2.5 | 210 ± 19 |
| 65 | -49.5 ± 2.1 | 380 ± 10 | 87.7 ± 2.3 | 215 ± 19 |
| 75 | -50.4 ± 2.2 | 397 ± 10 | 83.9 ± 2.1 | 213 ± 19 |
| 85 | -51.3 ± 2.1 | 405 ± 9 | 82.3 ± 1.8 | 216 ± 18 |
| 95 | -54.1 ± 2.1 | 414 ± 9 | 80.5 ± 1.7 | 236 ± 19 |

Altogether, the wire and thin film test combined show that this setup is capable of measuring the Seebeck coefficient within a few μV/K and resistivity (conductivity) with an uncertainty of a few percent for the film and ~10% for the wire. These uncertainties are conservatively estimated in that they are mostly based on the standard deviation of the raw data measured every 0.5 s. If we consider the averaging of entire data points (100 points in the current test), the uncertainty of the averaged values should be much smaller.

## V. CONCLUSION

In this study, we developed an affordable and easy-to-use thermoelectric measurement setup using Arduino and LabVIEW that can measure both the Seebeck coefficient and power factor simultaneously. By combining amplifiers and noise reduction schemes with the Arduino board, we were able to improve the resolution by over 1000 times, which thus allows measurement of sensitive thermoelectric signals. We tested this setup on a Chromel wire, which has well-defined Seebeck coefficient and resistivity values, and confirmed that our measurements are consistent with the literature values to within ~10%. We then measured a $Bi_2Se_3$ thin film, and confirmed that its thermoelectric properties are also consistent with previous results. In other words, this setup can provide quantitatively reliable thermoelectric measurements (with an uncertainty on the level of a few μV/K for Seebeck coefficients) even without sophisticated research-grade equipment and complex electrical contact schemes. Moreover, considering that this setup is compatible with wires as well as any slab-geometry samples, regardless of their being thin films or bulk materials, one should be able to use this setup for a variety of materials. Altogether, this thermoelectric measurement setup has lowered the barrier to thermoelectric studies both financially and technically. Thus, experimental studies of thermoelectricity are easily accessible not only to students but also to individual researchers with limited resources and experience.

## SUPPLEMENTARY MATERIAL

A link to the LabVIEW program will be provided here.

## ACKNOWLEDGMENTS

We would like to thank the anonymous reviewers for their valuable input, which significantly helped improve the paper. We would also like to thank the Monmouth County Board of Commissioners, for their support of MCVSD, the MCVSD Board of Education and Administration, including Dr. Charles Ford and Mr. Sean Meehan, the HTHS faculty and administrators, including Ms. Teresa Hough for her support of the research program, and the PFA. C.J.S. and S.O. are supported by National Science Foundation's DMR 2451900.

## AUTHOR DECLARATIONS

### Conflict of Interest

The authors have no conflicts to disclose.

### Author Contributions

A.J.O. built the setup, made the LabVIEW program, and carried out the measurement and data analysis under the guidance of C.Q. and S.O., as part of a research course taught by C.Q. C.J.S. synthesized the $Bi_2Se_3$ sample, under the guidance of S.O. A.J.O. wrote the initial draft under the guidance of C.Q. A.J.O. and S.O. edited and finalized the paper.

## APPENDIX

**TABLE A. List of parts used for the setup.** A breadboard, a DC power supply, jumper wires, cables, alligator clips, Kapton tape, resistors, a transistor heatsink, a rubber foot, a weight (~250 g) and a PC were also used but not listed in the table below.

| **Name/Description** | **Model** | **Vendor/Store** | **Quantity used** | **Price as of 09/2025** |
|---|---|---|---|---|
| Arduino Uno R3 compatible board with accessories | | Amazon | 1 | $8.99 for a piece |
| Thermoelectric cooler | TEC1-12706 | Amazon | 2 | $8.99 for 2 pieces |
| AD8495 thermocouple amplifier breakout board | | Adafruit | 3 | $11.95 for a piece |
| Insulated K-type thermocouple, AWG 30 | TC-NFP-1-6m | Minnesota Measurement Instruments | 1 ft | $4.99 for 1 m |
| PWM-to-Analog converter | | EC Buying/Amazon | 2 | $6.99 for a piece |
| N-channel MOSFET transistor | RFP30N06LE | Amazon | 3 | $5.99 for 12 pieces |
| Conducting copper tape, 1" width | | Bomei Pack/Amazon | 1 ft | $9.99 for 66 ft |
| Heatsink for the thermoelectric cooler | 404010B10-8810 | Easycargo/Amazon | 2 | $6.99 for 4 pieces |
| Thermal adhesive tape, 10 mm width | | Amazon | 1 ft | $5.89 for 25 m |